\documentclass[twocolumn,preprintnumbers,superscriptaddress,amsmath,amssymb,longbibliography]{revtex4-2}

\usepackage{graphicx}
\usepackage{dcolumn}
\usepackage{float} 
\usepackage{bbm}
\usepackage[dvipsnames]{xcolor}
\usepackage[colorlinks,linkcolor=blue,urlcolor=blue,citecolor=blue,anchorcolor=blue]{hyperref}
\usepackage{setspace}
\usepackage{xpatch} 
\usepackage{soul}

\usepackage[dvipsnames]{xcolor}
\def \markColorOne {Black}

\newcommand{\beginsupplement}{%
    \setcounter{section}{0}
    \renewcommand{\thesection}{S.\arabic{section}}
    \setcounter{figure}{0}
    \renewcommand{\thefigure}{S\arabic{figure}}
    \setcounter{table}{0}
    \renewcommand{\thetable}{S\arabic{table}}
    
    \setcounter{equation}{0}
    \renewcommand{\theequation}{S\arabic{equation}}
}

\begin{document}
\title{Full-counting statistics and quantum information of dispersive readout with a squeezed environment}

\author{Ming Li}

\affiliation{Shenzhen Institute for Quantum Science and Engineering,
Southern University of Science and Technology, Shenzhen 518055, China}
\affiliation{Guangdong Provincial Key Laboratory of Quantum Science and Engineering, Southern University of Science and Technology, Shenzhen, 518055, China}

\author{JunYan Luo}

\affiliation{Department of Physics, Zhejiang University of Science and Technology, Hangzhou 310023, China}

\author{Gloria Platero}

\affiliation{Instituto de Ciencia de Materiales de Madrid ICMM-CSIC, 28049 Madrid, Spain}

\author{Georg Engelhardt}
\email{georg-engelhardt-research@outlook.com}
\affiliation{Shenzhen International Quantum Academy, Shenzhen 518048, China}

\date{\today}

\pacs{
  }

\begin{abstract}
Motivated by the importance of dispersive readout in quantum technology, we study a prototypical dispersive readout setup that is probed by a squeezed vacuum in a time-reversal-symmetric fashion. To this end, we develop a full-counting-statistics framework for dispersive readout and analyze its measurement information, accompanied by a generalized mean-field approach suitable to deal with non-unitary dynamics. Distinct from conventional input-output theory, our full-counting-statistics approach enables the direct calculation of arbitrary-order cumulants for the measured cumulative (i.e., time-integrated) photonic distribution while maintaining applicability to nonlinear systems. The corresponding Fisher information exhibits an exponential dependence on the squeezing parameter and a robustness against residual nonlinearity, which can even approach the quantum Fisher information, setting an upper limit. This work introduces a conceptually streamlined and computationally efficient framework for continuous quantum measurements, making it well suited for widespread adoption in quantum technologies.
\end{abstract}

\maketitle

\allowdisplaybreaks

\emph{Introduction.---}  Dispersive readout of qubits stands as a pivotal technique for achieving quantum information processing and quantum sensing \cite{blais_cavity_2004, walter_rapid_2017, krantz_quantum_2019,ripoll2022quantum}. By exploiting the state-dependent shift in the resonance frequency of a coupled electromagnetic resonator, this method enables repeated, minimally invasive interrogation of the quantum state \cite{wallraff_approaching_2005,krantz_single-shot_2016,  kohler_dispersive_2018, swiadek_enhancing_2024,peri2024unified}. It has been utilized across quantum simulation of many-body systems~\cite{PerezGonzalez2019,zhang_synthesizing_2022, mansikkamaki_beyond_2022,PerezGonzalez2022}, fault-tolerant quantum computing \cite{raussendorf_fault-tolerant_2007, barends_superconducting_2014}, and quantum-error-correction protocols \cite{krinner_realizing_2022, kelly_state_2015, schindler_experimental_2011}.  Injection squeezing engineering \cite{drummond_quantum_2004,lutkenhaus_mimicking_1998,breuer_theory_2007,scully_quantum_1997,walls_quantum_2008}, which strategically redistributes non-commutative uncertainties between conjugate quadratures, has emerged as a powerful technique in quantum readout technology, e.g., for gravitational-wave detection \cite{acernese_increasing_2019,tse_quantum-enhanced_2019, barsotti_squeezed_2019}, axion-dark-matter search \cite{Malnou_Optimal_2018, malnou_squeezed_2019} and quantum computing \cite{qin_exponentially_2024, kam_fast_2024}. However, despite its importance, the simultaneous achievement of fast measurement and high fidelity remains inherently challenging \cite{walterRapidHighFidelitySingleShot2017, schuster_ac_2005,eickbusch_fast_2022,kirchmair_observation_2013,gambetta_qubit-photon_2006,minev_catch_2019,dumas_measurement-induced_2024,Li2025}.

The input-output formalism remains a workhorse for quantifying dispersive readout in quantum systems~\cite{drummond_quantum_2004,gardiner_quantum_2010,walls_quantum_2008,ripoll2022quantum}, offering analytical efficiency in weakly coupled linear regimes. Nevertheless,  even in prototypical systems, this framework exhibits fundamental limitations when addressing nonlinear dynamics or extracting information encoded in higher-order statistical moments of the output field~\cite{blais_circuit_2021}.
This crucial deficiency precludes microscopic descriptions of system dynamics and obscures physics beyond linear response, necessitating theoretical frameworks that transcend conventional input-output approaches.

In this Letter, we combine the quantum trajectory formalism~\cite{wiseman_quantum_2009} with full-counting statistics  (FCS)~\cite{Levitov1996,Schoenhammer2007,Cerrillo2016,Niklas2016,Wang2017,Engelhardt2017,Schaller2018,landi_current_2024}, establishing a unified framework that is analytically tractable and computationally feasible. Originally devised for photon emission and electron transport characterization~\cite{mandel_sub-poissonian_1979, levitov_charge_1993}, FCS has been successfully extended to counting photons in coherent driving fields~\cite{engelhardt_photon-resolved_2024, engelhardt_photon-resolved_2024-1,landi_current_2024, brange_photon_2019, portugal_photon_2023, li2024scalable}, to quantify the thermal transport in heat engines, and in the context of phase transitions \cite{wang_distinguishing_2024,gomez-ruiz_full_2020,pallister_phase_2025}. This statistical approach provides deeper insights for probing non-Gaussian features in quantum states beyond conventional methods.

\begin{figure*}[t]
\center
\includegraphics[width=2\columnwidth]{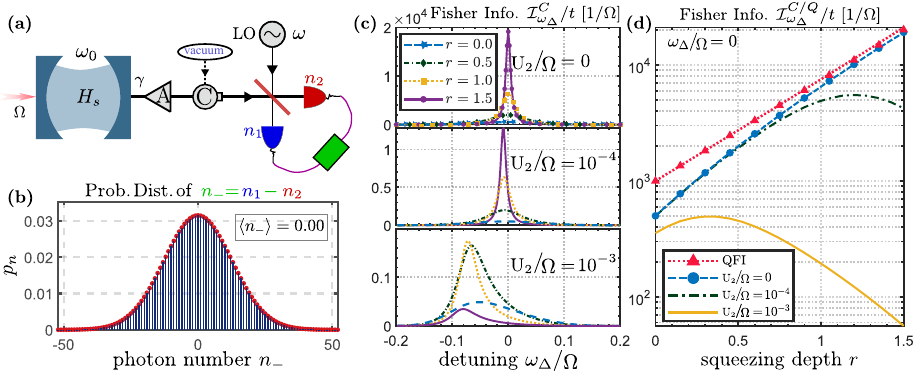}
\caption{ (a) Sketch of the dispersive-readout setup to estimate the frequency shift of the resonator, which is probed from the left port with Rabi frequency $\Omega$. Here, a phase-sensitive  amplifier  (A) is placed between the  circulator (C) and the resonator, such that the incoming vacuum field is both squeezed and antisqueezed before finally reaching the measurement setup, consisting of a beam splitter, a local oscillator (LO), and two photon detectors determining the photon number difference. (b) Probability distribution of the accumulated photon-number difference. (c) Cumulative Fisher information as a function of detuning for increasing nonlinearities $U_2=0, U_2=10^{-4}\Omega, U_2=10^{-3}\Omega$ depicted in the top, middle, and bottom panels, respectively.  The numerical (analytical) results for various squeezing strengths $r$, annotated in the legend, are depicted by dashed (pentagrams), dashed-dotted (rhombi), dotted (squares), and solid (circles) lines.
(d) Cumulative and quantum Fisher information as a function of squeezing depth for $\omega_\Delta=0$. The cumulative Fisher information [blue dashed line (numerics), circle markers (analytics)] increases exponentially and approaches the quantum Fisher Information [red dashed line (numerics), triangle markers (analytics)] as the squeezing depth increases. The cumulative Fisher information in the presence of non-linearity [green (dashed-dotted) and yellow (solid)] exhibits a turnover. Overall parameters are $\beta^2=10\Omega$, $\gamma=0.4\Omega$.}
\label{figSystem}
\end{figure*}

 We investigate an experimentally feasible dispersive-readout setup featuring a squeezed environment~\cite{backes_quantum_2021, Malnou_Optimal_2018, malnou_squeezed_2019,didier_perfect_2014, kronwald_arbitrarily_2013,quijandria2013circuit, Sanchez_Squeezed_2021, Navarrete_Inducing_2014}  embedded in a time-reversal-symmetric measurement setup~\cite{acernese_increasing_2019,tse_quantum-enhanced_2019, barsotti_squeezed_2019,Malnou_Optimal_2018, malnou_squeezed_2019, Malnou_Optimal_2018, malnou_squeezed_2019,backes_quantum_2021, wang2024quantum,kamble2024quantum} as shown in Fig.~\ref{figSystem}(a)-(d). Such a setup has been used in the search of axion dark matter and gravitational waves~\cite{acernese_increasing_2019,tse_quantum-enhanced_2019, barsotti_squeezed_2019,Malnou_Optimal_2018, malnou_squeezed_2019}. Using our FCS framework, we analytically demonstrate that squeezed injection induces a dramatic exponential enhancement of the Fisher information of the cumulative (i.e., time-integrated) photonic probability distribution,  which is similar to the findings in a non-time-reversal symmetric configuration~\cite{qin_exponentially_2024}. Remarkably, our methods reveal that this enhancement exhibits quantum optimality in the strong squeezing regime, by approaching the quantum Fisher information, which sets an upper limit on the measurement precision via the quantum Cramer-Rao bound~\cite{helstrom1967minimum,Hayashi2017}. This eliminates the need to deploy sophisticated, yet technically challanging means such as quantum decoders~\cite{yang_efficient_2023}.   Furthermore, we investigate the behavior of higher-order cumulants under varying squeezing strengths. We also study the Fisher information under residual nonlinearity \cite{boissonneault_dispersive_2009,Wang_Ideal_2019,dassonneville2020fast,mori2025high}, demonstrating robustness against small nonlinear effects, as compared to the probe amplitude. Conversely, for large nonlinearity, squeezing fails to enhance dispersive readout.

\emph{System.---} {\color{\markColorOne} Dispersive readout is a common detection method used in cavity-QED for the readout of qubit states~\cite{blais_circuit_2021}.  The qubit is thereby off-resonantly coupled to a microwave resonator, leading to a qubit-state-dependent shift of the resonator frequency. As shown in the setup in Fig.~\ref{figSystem}(a), a monochromatic probe field entering from the left port drives the resonator, leading to the emission of an output field at the right port, whose phase, which can be measured using the homodyne detection setup, depends on the frequency shift of the resonator.  }
The microscopic Hamiltonian of the dispersive readout setup in Fig.~\ref{figSystem}(a)  can be described by
\begin{eqnarray}
\hat H  &=& \hat H_{S}+\sum_{l,k }  \omega_{l,k}  \hat b_{l,k}^\dagger\hat b_{l,k} \nonumber\\
&&+\sum_{l,k}  g_{l,k} \left(  \hat a  - \hat a^\dagger  \right)  \left( \hat b_{l,k}^\dagger   -   \hat b_{l,k} \right),
\label{phy_model}
\end{eqnarray}
where $\hat{H}_{S} =\omega_0  \hat{a}^\dagger\hat{a} + \hat H_{NL}$ describes a resonator with frequency $\omega_0$, and Kerr nonlinearity $\hat H_{NL} = \frac{U_{2}}{2} (\hat{a}^\dagger)^2 \hat{a}^2$ with strength $U_2$. {\color{\markColorOne}  For simplicity, we neglect the qubit subsystem, but note that the frequency $\omega_0$ implicitly depends on the qubit state. For this reason, we investigate here the dispersive readout efficiency in terms of the estimation sensitivity of $\omega_0$, which can be quantified by the (classical) Fisher information as explained below.  Both the qubit-state dependent $\omega_0$ and the emergent Kerr nonlinearity arise from a Schrieffer-Wolff transformation as shown in the Supplementary Information~\cite{Boissonneault_Nonlinear_2008, boissonneault_dispersive_2009, Wang_Ideal_2019,blais_circuit_2021}.}

The index $l =\text{L},\text{R}$ refers to the transmission lines that are attached to the left and right resonator ports, respectively. Photonic modes are labeled $l, k$ and have energies $\omega_{l,k}$, which determine the propagation of electromagnetic signals in the transmission lines. The term in the second line describes the coupling between the resonator and the photonic transmission line operators $\hat{b}_{l,k}$.  { \color{\markColorOne} To measure the phase of the output field entering the right transmission line, the output field is mixed with a strong local oscillator $\beta$  and measured by two detectors $j=1,2$, to determine the time-integrated photon-number $\hat n_j$. The difference $\hat n_-= \hat n_1 -\hat n_2 $ is proportional to the quadrature $\hat n_- \propto  \hat{X} = \beta\hat{a} + \beta^*\hat{a}^{\dagger} \propto \Delta \varphi $ (see Supplemental Information) which is typically used to investigate the dispersive curve (i.e., the phase shift $\Delta \varphi$ as a function of the resonator frequency). }

We propose engineering the resonator environment by applying a phase-sensitive amplifier (A)  between the right output port (R)  and the circulator (C) as sketched in Fig.~\ref{figSystem}(a), which squeezes the transmission line modes as 
\begin{equation}
\hat b_{R,k }  \rightarrow   u   \hat b_{R,k}  - v   \hat  b_{R,k}^\dagger,
\end{equation}
with $u=\cosh r$ and $v=\sinh r$, where $r$ parameterizes the squeezing strength. Motivated by previous investigations of quantum metrology in optical systems~\cite{kamble2024quantum,wang2024quantum}, the squeezing operation is located between the resonator and the circulator in Fig.~\ref{figSystem}(a), such that its overall action exhibits time-reversal symmetry, i.e., the squeezing of the outgoing field perfectly reverses the squeezing of the incoming field for an ideal reflection at the resonator (see also  Fig.~S.1 in the Supplementary Information).

\emph{FCS.---} We are interested in the probability distribution of the cumulative (i.e., time-integrated) photon numbers $\boldsymbol{n} = (n_1, n_2)$  measured at detector $j=1,2$, from which we intend to estimate $\omega_0$. The corresponding moment-generating function  is defined as the inverse Fourier transform of the photon-number probability distribution,
\begin{equation}
M_{\boldsymbol{\chi}} := \sum_{\boldsymbol{n}} p_{\boldsymbol{n}} e^{-i\boldsymbol{\chi} \cdot \boldsymbol{n}},
\label{eq:momentGenFct}
\end{equation}
where {\color{\markColorOne} $\boldsymbol{\chi} = (\chi_1, \chi_2)$ is the vector of the scalar counting fields conjugated to $\boldsymbol{n}$,  enabling the characterization of the photon statistics at the two detectors.} As elaborated in the Supplementary Information, the moment-generating function can be expressed as $M_{\boldsymbol{\chi}} = \textrm{Tr}[\rho_{\boldsymbol{\chi}}]$, where $\rho_{\boldsymbol{\chi}}$ is the generalized reduced density matrix of the  resonator system, governed by the generalized master equation
\begin{eqnarray}
\frac{d}{dt}\hat\rho_{\boldsymbol{\chi}} &=& -i \left[\omega_\Delta  \hat{a}^\dagger\hat{a} + H_{NL}+\hat{H}_p, \hat\rho_{\boldsymbol{\chi}} \right] \nonumber \\
&& +\frac{1}{2} D_{\chi_1} \left[ \sqrt{\gamma} u\hat{a} - \sqrt{\gamma}v\hat{a}^\dagger - i \beta \right] \hat\rho_{\boldsymbol{\chi}} \nonumber \\
&& +\frac{1}{2} D_{\chi_2} \left[ \sqrt{\gamma} u\hat{a} - \sqrt{\gamma}v\hat{a}^\dagger + i \beta \right] \hat\rho_{\boldsymbol{\chi}}.
\label{generalized_master_fcs}
\end{eqnarray}
Here, $\omega_\Delta = \omega_0-\omega$ denotes the detuning between the resonator mode and the local oscillator mode, $\gamma$ is the effective dissipation rate due to the coupling to the transmission lines,  and $H_p = i \Omega \hat a + \text{h.c.}$ is the probe-field Hamiltonian. The information about the squeezing enters in the generalized dissipators $\hat{\mathcal{D}}_{\chi}[\hat{A}]\cdot = e^{-i\chi}\hat{A}\cdot\hat{A}^{\dagger} - \tfrac{1}{2}\{\hat{A}^{\dagger}\hat{A}, \cdot\}$.

In the following, we will analyze the probability distribution of the photon-number difference $n_- =n_1 -n_2$, which can be retrieved from $M_{\boldsymbol{\chi}}$ by a Fourier transformation for $\chi_1=-\chi_2 =\chi$,  and is shown in Fig.~\ref{figSystem}(b). The corresponding cumulants are systematically extracted via derivatives of the cumulant-generating function $K_{\chi}:= \log M_{\chi}$, with the $\ell$th-order cumulant  given by
\begin{equation}
\kappa_{\ell}(t) = \left. \frac{d^{\ell}}{d(-i\chi)^{\ell}} K_{\chi}(t) \right|_{\boldsymbol{\chi} = \boldsymbol{0}}.
\label{eq:def_cumulantsGeneral}
\end{equation}
Here, $\kappa_1$ gives the mean photon number and $\kappa_2$ the variance, while higher-order cumulants characterize non-Gaussian properties: $\kappa_3$ (skewness) quantifies distributional asymmetry, $\kappa_4$ (kurtosis) measures tail weight, and cumulants of order $\ell \geq 5$ capture increasingly subtle deviations from the  Gaussian statistics.

\textit{Non-Hermitian mean-field theory}.--- The standard mean-field approach transforms the annihilation operator into a  complex scalar $\alpha$ plus a small fluctuation operator $\delta \hat{a}$, expressed as $\hat{a} = \alpha + \delta \hat{a}$. This method is widely used in semiclassical approximations by setting the first-order fluctuations to zero. However,  the presence of the counting field renders the master equation in Eq.~\eqref{generalized_master_fcs}  non-Hermitian. Consequently, the standard mean-field approach fails because the first-order fluctuation terms differ depending on their position relative to the density matrix and the photonic operators. To address this, we introduce a generalized mean-field approach for our non-Hermitian dynamics by defining four distinct mean fields $\alpha_{\text{f}},\alpha_{\text{f}}^+,\alpha_{\text{b}},\alpha_{\text{b}}^+$ based on the relative ordering of the operators with respect to the density matrix, i.e., acting from the front (f) or back (b),
\begin{eqnarray}
\hat{a}\hat\rho_{\boldsymbol{\chi}}=\delta \hat{a}\hat\rho_{\boldsymbol{\chi}} + \alpha_{\text{f}}\hat\rho_{\boldsymbol{\chi}},&& \quad \hat\rho_{\boldsymbol{\chi}}\hat{a}=\hat\rho_{\boldsymbol{\chi}}\delta \hat{a} + \alpha_{\text{b}} \hat\rho_{\boldsymbol{\chi}},\nonumber\\
\hat{a}^\dagger \hat\rho_{\boldsymbol{\chi}}=\delta \hat{a}^\dagger \hat\rho_{\boldsymbol{\chi}} + \alpha_{\text{f}}^+\hat\rho_{\boldsymbol{\chi}}, && \quad \hat\rho_{\boldsymbol{\chi}}\hat{a}^\dagger=\hat\rho_{\boldsymbol{\chi}}\delta \hat{a}^\dagger + \alpha_{\text{b}}^+ \hat\rho_{\boldsymbol{\chi}}.
\label{NonHermitianMeanfield}
 \end{eqnarray}
This leads to the master equation
\begin{equation}
\frac{d}{dt}\hat\rho_{\boldsymbol{\chi}} =\mathcal{K}_{\boldsymbol{\chi}}\hat\rho_{\boldsymbol{\chi}} +  \mathcal{L}_{\boldsymbol{\chi}}^{(1)}[\delta \hat{a}]\hat\rho_{\boldsymbol{\chi}} + \mathcal{L}_{\boldsymbol{\chi}}^{(2)}[\delta \hat{a}]\hat\rho_{\boldsymbol{\chi}}  ,
\end{equation}
where $\mathcal{L}_{\boldsymbol{\chi}}^{(1)}[\delta \hat{a}]\hat\rho_{\boldsymbol{\chi}} \sim \mathcal{O}(\delta \hat{a})$ captures the linear fluctuation dependence, while $\mathcal{L}_{\boldsymbol{\chi}}^{(2)}[\delta \hat{a}]\hat\rho_{\boldsymbol{\chi}} \sim \mathcal{O}(\delta \hat{a}^2)$ contains all higher-order fluctuations. By selecting appropriate mean fields ($\alpha_{\text{f}}$, $\alpha_{\text{f}}^+$, $\alpha_{\text{b}}$, $\alpha_{\text{b}}^+$) that satisfy $\mathcal{L}_{\boldsymbol{\chi}}^{(1)}[\delta \hat{a}]\hat\rho_{\boldsymbol{\chi}} = 0$ and neglecting higher-order terms, we eventually find that the moment-generating function $M_{\boldsymbol{\chi}}$ fulfills
\begin{equation}
\frac{d}{dt}M_{\boldsymbol{\chi}} = \mathcal{K}_{\boldsymbol{\chi}}\mathcal{M}_{\boldsymbol{\chi}},
\end{equation}
such that the cumulant-generating function scales linearly in time  $K_{\boldsymbol{\chi}}(t\rightarrow \infty) = \mathcal{K}_{\boldsymbol{\chi}} t$, with the coefficient given by
\begin{eqnarray}
\mathcal K_{\boldsymbol{\chi}}   = -\frac{2 \sin^2(\frac{\chi}{2}) \left[ 16 \beta^2 \omega_\Delta^4 + f_3 \omega_\Delta^3 + f_2\omega_\Delta^2   + f_1 \omega_\Delta + f_0 \right]}{16 \omega_\Delta^2 \gamma^2 \sinh^2(2r) \sin^2 \chi + (4\omega_\Delta^2 + \gamma^2)^2},\nonumber \\
\end{eqnarray}
and
\begin{eqnarray}
f_3&=&-32 i\beta \sqrt{\gamma} \Omega e^r \cot\frac{\chi}{2},\nonumber\\
f_2&=& 8 \beta^2 \gamma^2 (e^{4r} - 1) \cos\chi  + 8 \beta^2 \gamma^2 e^{4r} + 16 \gamma \Omega^2 e^{2r},\nonumber\\
f_1&=&-8i\beta\gamma^{5/2}\Omega(e^r \cos\chi\cot\frac{\chi}{2}+e^{-3r}\sin\chi),\nonumber\\
f_0&=& \beta^2 \gamma^4 + 4 \gamma^3 \Omega^2 e^{-2r} ,
\end{eqnarray}
which can be used to obtain the cumulants of arbitrary order via Eq.~\eqref{eq:def_cumulantsGeneral}. The linear time-dependence of the cumulant-generating function reflects the cumulative character of the measurement process.

\emph{Fisher information---} 
{ \color{\markColorOne} Here we investigate the measurement sensitivity for estimating $\omega_{\Delta}$, as this is directly related to the qubit-state readout.} According to the quantum-Cramer-Rao bound~\cite{helstrom1967minimum,Hayashi2017}, the sensitivity is bounded by $\left<\hat \omega_\Delta^2  \right>\geq 1/\mathcal{I}_{\omega_\Delta}^{\text{C}} \geq 1/\mathcal{I}_{\omega_\Delta}^{\text{Q}}$, where $\hat \omega_\Delta$ denotes an arbitrary estimator for $\omega_\Delta$~\cite{Gorecki2025}. The cumulative Fisher information (CFI) $\mathcal{I}_{\omega_\Delta}^{\text{C}}$ specifies the precision limit in terms of  the probability distribution $p_{n_-}$ for the cumulative photon number difference $n_-$. For long measurement times $t\rightarrow \infty$, this distribution approaches a Gaussian function, such that the CFI,
\begin{equation}
{\cal I}_{\omega_\Delta}^{\text{C}} \equiv \sum_{n_-} p_{n_-}  \left(\frac{\partial}{\partial \omega_\Delta}\textrm{ln}p_{n_-}  \right)^2 \rightarrow   \frac{\left(\partial\kappa_1/\partial\omega_{\Delta} \right)^2}{\kappa_2}  t , 
\end{equation}
can be evaluated in terms of the first two cumulants.  The quantum Fisher information (QFI) $\mathcal{I}_{\omega_\Delta}^{\text{Q}}$ quantifies the ultimate measurement precision by an arbitrary projective measurement of the quantum state of the total system (resonator and transmission lines), and can be evaluated by integration of a generalized master equation~\cite{gammelmark2014fisher}. 

\emph{Linear system.---} In the absence of nonlinearities we find that the CFI approaches  
\begin{equation}
{\cal I}_{\omega_\Delta}^{\text{C}} = \frac{64 \beta^2 \gamma \left(\gamma^2 - 4 \omega_\Delta^2\right)^2 \Omega^2 e^{2 r} }{\left(4 \omega_\Delta^2 + \gamma^2\right)^2 \Big[ \gamma^4\beta^2  + 4 \Omega^2 \gamma^3  e^{-2 r}  + \omega_\Delta^2\Phi\Big]} \,t,
\label{eq:Fisher_full}
\end{equation}
where $\Phi=16 \beta^2 \omega_\Delta^2 +8\beta^2 \gamma^2 (2 e^{4 r} - 1) + 16 \gamma \Omega^2 e^{2 r} $ as shown in the Supplementary Information, which achieves its global maximum  at resonance $\omega_\Delta=0$. Interestingly, increasing the squeezing depth $r$ sharpens the peak around resonance, and thus reduces the measurement bandwidth.  At zero detuning $\omega_\Delta=0$,
\begin{equation}
{\cal I}_{\omega_\Delta}^{\text{C}} = \frac{64 \beta^2 \Omega^2 e^{2 r} }{ \gamma^3\beta^2  + 4 \Omega^2 \gamma^2  e^{-2 r} } \, t \approx  \frac{64  \Omega^2  }{ \gamma^3 } \, e^{2 r}  \, t  \quad (r \gg 1),
\label{eq:Fisher_res}
\end{equation}
revealing an exponential quantum enhancement with ${\cal I}_{\omega_\Delta}^{\text{C}}  \sim e^{2r} $. Remarkably, the numerical benchmarking in Fig.~\ref{figSystem}(d)  exactly agrees with the non-Hermitian mean-field approach.

As shown in the Supplementary Information using the non-Hermitian mean-field approach, the QFI in the asymptotic limit becomes
\begin{equation}
{\cal I}_{\omega_\Delta}^{\text{Q}} = \frac{64 \Omega^2 \gamma }{\left(4 \omega_\Delta ^2 + \gamma^2\right)^2}\, e^{2 r}\,t.
\label{eq:QFIAnalytical}
\end{equation}
Crucially, Eq.~\eqref{eq:Fisher_full} and \eqref{eq:QFIAnalytical} fulfill ${\cal I}_{\omega_\Delta}^{\text{Q}} > {\cal I}_{\omega_\Delta}^{\text{C}}$ for all parameters, reflecting the microscopic consistency of the applied quantum trajectory approach amended with FCS. To quantify the efficiency of the cumulative measurement statistics, we introduce the quantum efficiency  $\eta \equiv \mathcal{I}_{\omega_\Delta}^{\text{C}} /  {\mathcal{I}^{Q}_{\omega_\Delta}}$.
At zero detuning $\omega_\Delta=0$, this simplifies to $\eta = 1 - \frac{4\Omega^2}{4\Omega^2+\gamma \beta^2 e^{2r}}$, revealing a surprising insight. As the squeezing strength $r$ increases,  $\eta \to 1$, i.e., the cumulative measurement statistics approaches quantum optimality set by the quantum Cramer-Rao bound. Moreover, independent of the squeezing strength $r$, the quantum efficiency of the cumulative measurement protocol is never less than 50\% of the quantum Fisher information. {\color{\markColorOne} Finally, we note that the shot-noise limit in our model corresponds to the unsqueezed case r = 0, where the reservoir reduces to the common photonic vacuum.}

\begin{figure}[t]
\center
\includegraphics[width=1.0\columnwidth]{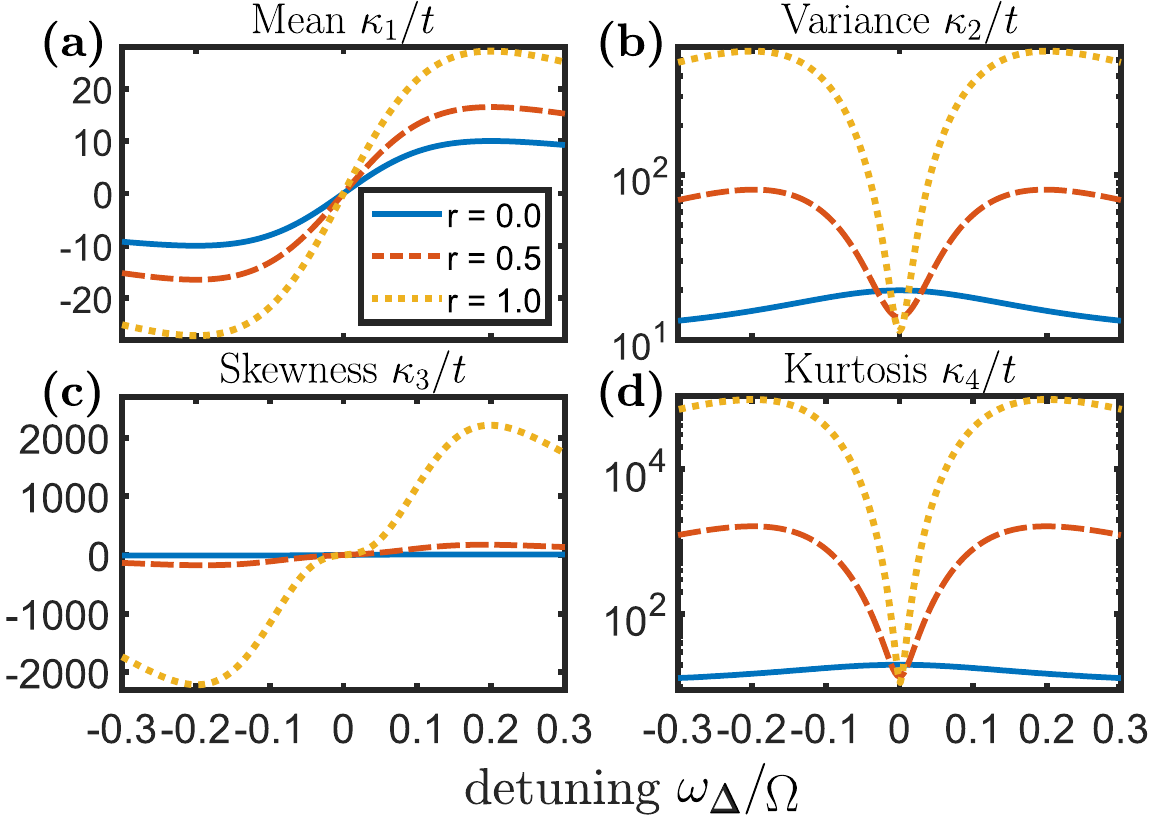}
\caption{First four cumulants as a function of detuning for different squeezing strenghts. Overall parameters are the same as in Fig.~\ref{figSystem}.}
\label{cumulants}
\end{figure}

To reveal the origin of the exponential enhancement of the Fisher information with squeezing strength and the quantum optimality, we analyze the first four cumulants as a function of detuning in Fig.~\ref{cumulants} for different squeezing strengths.  Near resonance, squeezing simultaneously enhances the gradient $\partial_{\omega_\Delta} \kappa_1$ [in  Fig.~\ref{cumulants}(a)], while suppressing the noise $\kappa_2$ in the measurement statistics [Fig.~\ref{cumulants}(b)]. This dual action creates the observed quantum-enhanced precision of the cumulative Fisher information. However, this advantage quickly disappears for a finite detuning. {\color{\markColorOne} As the local oscillator is not squeezed,  the detection shot-noise floor remains constant. The observed variance reduction stems from the squeezed reservoir acting on the resonator: the Bogoliubov-dressed dissipators in Eq.~(\ref{generalized_master_fcs}), $\mathcal{D}_{\chi_j}\!\bigl[\sqrt{\gamma}\,u\,\hat a - \sqrt{\gamma}\,v\,\hat a^\dagger \pm i\beta\bigr]$, induce intracavity quadrature squeezing that is revealed through the homodyne measurement.}

The high quantum efficiency at resonance is related to the Gaussian character of the cumulative measurement statistics. As we observe in Fig.~\ref{cumulants}(c) and Fig.~\ref{cumulants}(d), the third and the forth cumulants gradually vanish for increasing squeezing strength $r$ at $\omega_0=0$, rendering the probability distribution more Gaussian. It is known that for Gaussian probability distributions, the cumulative random variables (here the integrated photon number difference) constitute a sufficient statistics, i.e., contain all information about the relevant parameters in the probability distribution~\cite{kay1998fundamentals}. Consequently,  the first two cumulants fully characterize the probability distribution in the strong squeezing regime.

\begin{figure}[t]
\center
\includegraphics[width=1.\columnwidth]{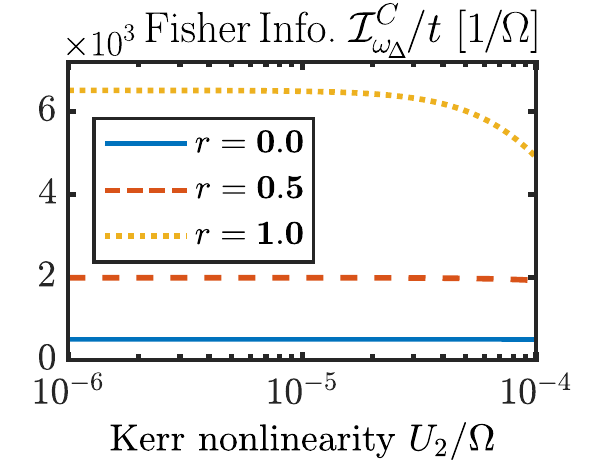}
\caption{Fisher information versus Kerr nonlinearity $U_2$. Overall parameters are the same as in Fig.~\ref{figSystem}.}
\label{FI_Nonlinearity}
\end{figure}

\emph{Nonlinearity.---}  Here we assess the impact of a finite Kerr nonlinearlity $U_2>0$. In Fig.~\ref{figSystem}(c), we observe that the symmetric shape of the CFI gradually disappears with stronger nonlinearities, creating configurations where lower squeezing outperforms higher squeezing.  Accordingly, we observe in Fig.~\ref{figSystem}(d)  that a finite nonlinearity leads to a turnover of the Fisher information as a function of the squeezing depth. Figure~\ref{FI_Nonlinearity} reveals that the CFI remains robust under weak nonlinear perturbations, i.e., it is constant as a function of $U_2$. However, for larger squeezing strengths, the cumulative Fisher information becomes increasingly susceptible to such nonlinearities, eventually degrading the precision advantage.

\emph{Conclusions.---}  
In this work, we applied a quantum-trajectory approach amended with FCS to systematically analyze the quantum information of dispersive readout in a squeezed environment. To facilitate the analysis, we
developed a non-Hermitian mean-field theory valid in the long-time limit.
This flexible approach also provides deeper insights into system dynamics through higher-order cumulants of the cumulative photonic probability distribution, the corresponding Fisher information [cumulative Fisher information (CFI)], and the quantum Fisher information (QFI). 

This analysis reveals that both the CFI and the QFI  exhibit exponential enhancement as a function of the squeezing parameter. Even more interesting, the CFI approaches the QFI in the large squeezing regime, signaling asymptotic quantum optimality fulfilling the quantum-Cramer-Rao bound. Moreover, we have demonstrated that this scaling is robust with respect to weak Kerr nonlinearities. {\color{\markColorOne} While the Kerr-nonlinearity has the potential to squeeze the resonator state, our microscopic treatment shows that it has, in fact, a detrimental effect on the Fisher information~\cite{Wang_Ideal_2019, boissonneault_dispersive_2009}. This contrasts previous investigations simply postulating linearized Kerr-nonlinearites to reach a sensitivity enhancement~\cite{Laflamme_Weak_2012}.}   As the methods developed in this work overcome limitations inherent in the celebrated input-output theory, they are suitable for the analysis of more complex resonator systems, including those exhibiting strong nonlinearity~\cite{puel2024confined,perezGonzalez2025lightmatter}, to investigate the emergence of novel features as discovered here. 

\textit{Acknowledgments.---}   J.Y.L. acknowledges the support from the National Natural Science Foundation of China (Grant Nos. 11774311). GP is supported by the Spanish Ministry of Science through the grant: PID2023-149072NBI00 and  by the CSIC Research Platform PTI-001.G.E. acknowledges the support by the National Natural Science Foundation of China (Grant No. W2432004).

\bibliography{bibliography}

\end{document}


\title{Full-counting statistics and quantum information of dispersive readout with a squeezed environment: Supplementary Materials}

\date{\today}

\author{Ming Li}

\affiliation{Shenzhen Institute for Quantum Science and Engineering,
Southern University of Science and Technology, Shenzhen 518055, China}
\affiliation{Guangdong Provincial Key Laboratory of Quantum Science and Engineering, Southern University of Science and Technology, Shenzhen, 518055, China}

\author{JunYan Luo}

\affiliation{Department of Physics, Zhejiang University of Science and Technology, Hangzhou 310023, China}

\author{Gloria Platero}

\affiliation{Instituto de Ciencia de Materiales de Madrid ICMM-CSIC, 28049 Madrid, Spain}

\author{Georg Engelhardt}
\email{georg-engelhardt-research@outlook.com}
\affiliation{International Quantum Academy, Shenzhen 518048, China}

\maketitle

\allowdisplaybreaks
\beginsupplement

\tableofcontents

\section{ Preliminaries}

\subsection{Effective dispersive Hamiltonian and its higher-order terms}

To connect the system Hamiltonian $H_S$ used in the main text with its microscopic origin, we briefly review how the dispersive Hamiltonian emerges from the Jaynes--Cummings (JC) model and list the leading higher-order corrections in the large-detuning regime following Refs.~\cite{Boissonneault_Nonlinear_2008, boissonneault_dispersive_2009, Wang_Ideal_2019}.

In a typical superconducting circuit QED setup, a two-level system (i.e., the qubit) couples transversely to a single resonator mode with strength $g_x$. In the large-detuning (dispersive) regime $g_x \ll |\Delta_{qr}|$, where $\Delta_{qr}= \omega_q-\omega_r$ and $\omega_q$ ($\omega_r$) is the qubit (resonator) frequency, the JC Hamiltonian reads
\begin{eqnarray}
\hat H = \omega_r \hat a^\dagger \hat a + \frac{\omega_q}{2}\hat\sigma_z
+ g_x\bigl(\hat a^\dagger \hat\sigma^- + \hat a\,\hat\sigma^+\bigr).
\label{eq:JC}
\end{eqnarray}
Here $\hat a$ ($\hat a^\dagger$) annihilates (creates) a resonator photon and $\hat\sigma^\pm$ are the qubit ladder operators.

Defining $\lambda = g_x/\Delta_{qr}$ and the total-excitation operator $\hat N_q = |e\rangle\langle e| + \hat a^\dagger \hat a$, one can exactly diagonalize Eq.~\eqref{eq:JC} by the unitary
\begin{eqnarray}
\hat U_{\rm dis} = \exp\!\Big[-\Lambda(\hat N_q)\,\bigl(\hat a^\dagger \hat\sigma^- - \hat a\,\hat\sigma^+\bigr)\Big],
\qquad
\Lambda(\hat N_q)= -\frac{\arctan\!\bigl(2\lambda\sqrt{\hat N_q}\bigr)}{2\lambda\sqrt{\hat N_q}}.
\end{eqnarray}
This yields an diagonal Hamiltonian (no series truncation is invoked),
\begin{eqnarray}
\hat H_{\rm diag} 
= \omega_r \hat a^\dagger \hat a + \frac{\omega_q}{2}\hat\sigma_z
- \frac{\Delta_{qr}}{2}\Bigl(1-\sqrt{1+4\lambda^2 \hat N_q}\Bigr)\hat\sigma_z .
\label{eq:diag}
\end{eqnarray}

To make the dispersive structure explicit, we expand the square root in Eq.~\eqref{eq:diag} for $|\lambda|\ll 1$ up to fourth order. Collecting terms of normal-ordered photon operators, one obtains the standard dispersive Hamiltonian with higher-order corrections,
\begin{eqnarray}
\hat H_{\rm dis}
&=& 
\bigl(\omega_r + \delta_r\bigr)\,\hat a^\dagger \hat a
+ \chi_z\, \hat a^\dagger \hat a\,\hat\sigma_z
+ \zeta\, \hat a^\dagger \hat a^\dagger \hat a \hat a\,\hat\sigma_z
+ \frac{1}{2}\bigl(\omega_q+\chi_z\bigr)\hat\sigma_z
+ \mathcal O(\lambda^6),
\label{eq:Hdisp}
\end{eqnarray}
where
\begin{eqnarray}
\chi_z = \frac{g_x^2}{\Delta_{qr}}\Bigl(1 - 2\lambda^2\Bigr) + \mathcal O(\lambda^4),\qquad
\delta_r \sim -\,\frac{g_x^4}{\Delta_{qr}^3} + \mathcal O(\lambda^6),\qquad
\zeta \sim -\,\frac{g_x^4}{\Delta_{qr}^3} + \mathcal O(\lambda^6)
\end{eqnarray}
denote, respectively, the leading dispersive (cross-Kerr) shift, the small state-independent fourth-order renormalization of the resonator frequency, and the qubit-state-dependent Kerr-scale correction.
Equation~\eqref{eq:Hdisp} shows that, conditional on the qubit state ($\hat\sigma_z=\pm 1$), the resonator frequency acquires a shift $\pm\chi_z$ and a small state-dependent self-Kerr proportional to $\zeta$.

In the main text we do not keep the qubit subspace. Instead, when the qubit is prepared in a definite computational state during the measurement, its state-dependent renormalization of the resonator frequency can be absorbed into the single-mode term, yielding
\begin{eqnarray}
\hat H_S = \omega_0 \hat a^\dagger \hat a + \hat H_{\rm NL},
\qquad
\hat H_{\rm NL} = \frac{U_2}{2}\,(\hat a^\dagger)^2 \hat a^2,
\end{eqnarray}
where $\omega_0=\omega_r \pm \chi_z + \mathcal O(\lambda^3)$ and the effective Kerr strength $U_2$ captures both residual circuit nonlinearity (e.g., from Josephson elements) and the dispersive-origin corrections (the $\zeta$-terms) collected into a compact quartic nonlinearity. This is precisely the $H_S$ used in Eq.~(1) of the main text and justifies treating the qubit’s influence as a state-conditioned renormalization of $\omega_0$ together with a weak Kerr nonlinearity that becomes relevant only beyond leading order in $g_x/\Delta_{qr}$.

\subsection{Relation between quadrature measurement and photon-counting}

Typically, homodyne measurement and dispersive readout consider the photonic quadrature operators of the resonator output field. This approach is very convenient when dealing with linear systems, allowing for exact analytical solutions. In contrast, we theoretically describe homodyne detection in terms of the measured photon number at the detectors. This is an equivalent approach, which becomes beneficial when dealing with nonlinear systems.

To see this equivalence, we consider the resonator output mode $\hat b_{0}$ and the local oscillator $\hat b_{L0}$. After mixing in a balanced beam splitter [see Fig. 1(a)], we obtain the output modes
\begin{eqnarray}
    \hat b_1 &=& \frac{1}{\sqrt{2}} \left( \hat b_{L0} +i\hat b_{0}   \right),\nonumber \\
    \hat b_2 &=& \frac{1}{\sqrt{2}} \left( \hat b_{L0} -i\hat b_{0}   \right).
\end{eqnarray}
Assuming that the local oscillator state is in a strong coherent state with $\hat b_{LO} = \beta e^{i\varphi}$ with real-valued $\beta$ and $\varphi$, we find that  the following relation for the photon-number-difference operator
\begin{eqnarray}
    \hat b_1^\dagger \hat b_1  - \hat b_2^\dagger \hat b_2 &\approx & \beta \left(\hat b_{0} e^{i\varphi} + \hat b_{0} e^{-i\varphi}  \right) ,
\end{eqnarray}
which is thus proportional to a quadrature of the output field. 

Here, we investigate the dispersive readout in terms of the photon number instead of the common photonic quadratures for the following three reasons:
\begin{itemize}
    \item Experimental setups typically measure intensities, which are more closely related to photon-number measurements than to the quadratures. The quadratures are thus only inferred from the intensity measurements.

    \item The description in terms of photon numbers allows us to take advantage of the rich methodologies of full-counting statistics. This makes it feasible to calculate higher-order cumulants and deal with nonlinear systems. In contrast, the traditional calculation of quadratures is relatively simple only when dealing with quadratic systems.

    \item We aim to evaluate the Fisher information of the measurement setup, which is defined in terms of a probability distribution. As the probability is discrete for a photon-number measurement, it is conceptually easier than the Fisher information of the continuous probability distribution of the photonic quadratures.
\end{itemize}

\section{Generalized Quantum Master Equation Incorporating Full Counting Statistics with Homodyne Detection}

\label{app:genMasterEquation}

In this section, we outline the derivation of the generalized master equation [Eq. (7)] from first principles by systematically incorporating full-counting statistics in the quantum trajectory formalism. This is  a complementary approach to the heuristic inclusion of the counting fields in Ref.~\cite{landi_current_2024}.

\subsection{Physical Setup}
The full system-reservoir model consists of two components: the quantum system and the external transmission lines. Their Hamiltonian includes both the free terms and the coupling terms, and is given by, 
\begin{eqnarray}
\hat H &=&\hat H_S + \hat H_{T}  +  \hat H_{ST},
\label{eq:hamiltonian:full}
\end{eqnarray}
where $\hat H_{S} =  \omega_0 \hat a^\dagger \hat a $ denotes the Hamiltonian of the resonator, which is quantized by the photonic operators $\hat a, \hat a^\dagger$. For the following derivations, we consider a linear resonator described by $\hat H_{S}$, while a Kerr nonlinearity is added phenomenologically at the end.

The Hamiltonian of the transmission lines is quantized by the photonic operators $\hat b_{l,k}$  and reads
\begin{eqnarray}
\hat H_{T} = \sum_{l,k}  \omega_{l,k}  \hat b_{l,k}^\dagger\hat b_{l,k} ,  
\label{eq:hamiltonianTransmissionLine}
\end{eqnarray}
where the index  $l =\text{L},\text{R}$ refers to the transmission lines which are attached from the left and right, respectively,  to the  resonator. The photonic modes are labeled by  $l, k$ and have energies   $   \omega_{l,k}$, which determine the propagation of electromagnetic signals within the transmission lines.

Finally, the coupling between the system and the transmission line is given by 
\begin{equation}
\hat H_{ST}	= \sum_{l,k}  g_l \left(  \hat a  - \hat a^\dagger  \right)  \left( \hat b_{l,k}^\dagger   -   \hat b_{l,k} \right).
\label{eq:ham:systemTransmissionLine}
\end{equation}

To calculate the moment-generating function we derive a generalized quantum master equation. To this end, we follow the quantum trajectory approach originally developed by Wiseman and Milburn~\cite{wiseman_quantum_2009}, which we represent as a generalized density matrix by including counting fields in the equation of motion~\cite{landi_current_2024}. In doing so, we can evaluate the moment-generating function on the ensemble level (i.e., the system density matrix), instead of simulating the stochastic dynamics of the unraveled quantum time evolution. 

\begin{figure*}[t]
\center
\includegraphics[width=0.65\columnwidth]{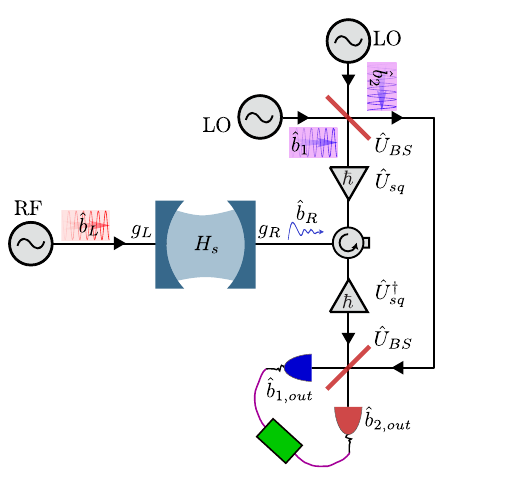}
\caption{ Schematic representation of the circuit used in the derivation of the generalized master equation, which is equivalent to the circuit in Fig. 1(a) in the Letter.}
\label{theofigSystem}
\end{figure*}

\subsection{Measurement basis and initial condition}

To accurately introduce the counting fields, we must represent the Hamiltonian in the basis of photonic operators,  in which we measure the system. We represent the transmission line modes with field operators as a function of position instead of frequency. To spare technical details, we give an informal introduction, and refer the reader to Ref.~\cite{Clerk2010} for an accurate introduction into the framework of windowed Fourier transformation. For brevity, we will carry out the following calculations for the transmission line $l=\text{R}$, while remarking that the same procedure also applies to the transmission line $l=\text{L}$.

Using the transmission-line operators in momentum space $\hat b_{\text{R},k}$, we distribute the transmission line $l=\text{R}$ in infinitesimal segments of length $dr$, which are labeled by $j$. For each line segment,  we define the directed field operators at position $r_j = j dr$ by
\begin{equation}
	\hat b_{\text{R}}(r_j)  = \sqrt{dr} \sum_{\left| k-K \right|<\epsilon_K }  e^{i(k-K) r_j} \hat b_{\text{R},k},
	\label{eq:localTransmissionLineOp}
\end{equation}
 where $K = \omega/c$ is a reference momentum determined by $\omega$, the frequency of the local osciallator. The parameter $\epsilon_K\ll K$ shall be of the order of magnitude of $g_{\text{R}}$, the coupling between the system and the transmission line. The normalization of the operators is chosen such that
\begin{equation}
	\left[ \hat b_{\text{R}}(r_{j_1}) ,\hat b_{\text{R}}^\dagger(r_{j_2})  \right] =  \delta_{j_1,j_2},
\end{equation}
i.e., a Kronecker delta instead of common delta function. This formally facilitates the introduction of counting fields, which count integer number of photons.

To make progress, we transform the Hamiltonian in Eq.~\eqref{eq:hamiltonian:full} into an interaction picture  defined by
\begin{equation}
\hat U_{0}(t) =e^{i\left( \hat H_S+ \hat H_{T} \right) t }.
\end{equation}
It is easy to see that
\begin{equation}
\hat U_{0}^\dagger (t) \hat 	a \hat U_{0}(t) =  \hat a  e^{-i \omega_0  t} .
\end{equation}
Moreover, the directed field operators fulfill
\begin{eqnarray}
\hat b_{\text{R}}(r_j,t)   
 &=& \hat  U_{0}^\dagger (t)	\hat b_{\text{R}}(r_j) \hat U_{0}^\dagger (t) \nonumber \\
&=& \hat b_{\text{R}}(r_j - ct) e^{-i\omega t},
\end{eqnarray}
i.e., they are translated along the positive direction of the tranmission line, while acquiring a dynamical phase determined by the reference frequency $\omega$.

The Hamiltonian in  the interaction picture in Eq.~\eqref{eq:hamiltonian:full} is thus  given by
\begin{eqnarray}
\hat H(t)  &=&  \sum_j g_j(t)   \left(\hat a e^{-i \omega_0 t}-  \hat a ^\dagger e^{i \omega_0 t}   \right)  \left(   \hat b_{\text{R},j}^\dagger  e^{i\omega t} -  \hat b_{\text{R},j} e^{-i\omega t}  \right) ,
\end{eqnarray}
where we have introduced the time-dependent system-transmission line couplings
\begin{eqnarray}
	g_j(t) &=&   \frac{ g_{\text{R}}  }{\sqrt{dr}} \Theta(r_{j+1} - ct ) \Theta(ct- r_j  )
\end{eqnarray}
with $\Theta(x)$ being the Heavy-side function. The time dependence of the couplings determines the instant in time in which the operator $\hat b_{\text{R},j} = \hat b_{\text{R}}(r_j)  $ interacts with the quantum system. Since the frequencies $\omega$ and $\omega_0$ are significantly larger than all other parameters in the system, we can apply the rotating-wave approximation (RWA),
\begin{eqnarray}
\hat H(t) 
&\approx&  \sum_j    g_j(t) \left( \hat a  \hat b_{R,j}^\dagger e^{-i \omega_\Delta  t}   +  \hat a^\dagger \hat b_{R,j}   e^{i \omega_\Delta t}    \right) ,
\label{eq:ham:rotatingWave}
\end{eqnarray}
where  $ \omega_{\Delta} = \omega_{0} -\omega $ is the detuning.

As previously stated, our objective is to probe the quantum system under the influence of a squeezed environment, followed by the measurement of the output light field after its interaction with the system, as illustrated in Fig. 1 in the Letter. We represent the unitary transformation associated with the beamsplitter as $\hat U_{BS}$ and the squeezing operation as $\hat U_{SQ}^\dagger$. 

For theoretical clarity, we use the symmetric representation in Fig.~\ref{theofigSystem} of the measurement circuit instead of the equivalent representation in Fig.~1(a) of the Letter. Such a symmetric representation enables the interpretation of the squeezing operations and the beam splitters as a basis transformation of the Schr\"{o}dinger equation, i.e.,
\begin{equation}
i\partial_t\left| \Psi_M(t) \right> = \hat H_{M}  \left| \Psi_M(t) \right>,
\end{equation}
where $\left| \Psi_M(t) \right> =\hat U_{SQ}^\dagger\hat U_{BS}^\dagger \left| \Psi(t) \right>$ and $\hat H_{M}=\hat U_{SQ}^\dagger\hat U_{BS}^\dagger\hat H\hat U_{SQ}\hat U_{BS}$. This transformation can be seen as a representation of the Hamiltonian in the basis, in which the system is measured (hence the index $M$). This representation is necessary for the introduction of the counting fields. 

Applying the basis transformation to the transmission-line operator in Eq.~\eqref{eq:localTransmissionLineOp}  yields
\begin{equation}
\hat b_{\text{R},j}  =  u  \left(  \hat b_{1,j} -  \hat b_{2,j} \right) - v \left(  \hat  b_{1,j}^\dagger  -  \hat b_{2,j}^\dagger \right),
\label{eq:measurementBasisTransformation}
\end{equation}
where $\hat b_{1,j}$ and $\hat b_{2,j}$ are the input operators in Fig.~\ref{theofigSystem}, which also represent the measurement basis of the homodyme measurement. To simplify the notation, we have absorbed the factor $\sqrt{2}$ into the parameters $u$ and $v$, and will restore them at the end by the replacement $u \to \frac{u}{\sqrt{2}}$ and $v \to \frac{v}{\sqrt{2}}$. To establish the counting-field formalism, we assume the initial state of the modes $\hat b_{\alpha,j}$ to be a product of coherent states
\begin{equation}
	\left| \phi_M(0) \right>  =  \bigotimes_{l=1,2;j} \left|\beta_{l,j} \right> ,
	\label{eq:theoreticalInitialCondition}
\end{equation}
where $
\hat b_{l,j}	\left|\beta_{l,j} \right>  = \beta_{l,j} \left|\beta_{l,j} \right> ,
\label{eq:fieldInitialState}
$
and we assume the scaling relation $\beta_{l,j} = \beta_{l} (dr/c)^{\frac{1}{2} }$. In doing so, we make sure that $\hat b_{l,j}$ satisfies
\begin{equation}
	\left<\beta_{l,j} \right| \hat b_{l,j}^\dagger \hat b_{l,j}  \left|\beta_{l,j} \right> = \left| \beta_{l}\right|^2 \frac{dr}{c},
\end{equation}
such that the mean photonic occupation number scales linearly with the infinitesimal length of the line segment. Crucially, the initial condition in Eq.~\eqref{eq:theoreticalInitialCondition} is a theoretical initial condition, which is distinct from the physical (i.e., the experimental) one. Denoting the actual physical initial condition by $\left| \phi(0) \right>$, we find the relation
\begin{equation}
\left| \phi(0) \right> =  \hat U_{BS} \left| \phi_M(0) \right> 
\label{eq:physicalTheorticalInitCondition}
\end{equation}
by comparing Fig.~\ref{theofigSystem} and Fig.~1, from which we can determine the theoretical amplitudes $\beta_{l,j}$. For instance, assuming a vacuum as experimental input field $\left| \phi(0) \right> =\left| vac \right>$, we find that
\begin{equation}
\beta_{1,j} =\beta_{2,j},
\end{equation} 
which gives rise to a destructive interference at the input-circuit beam splitter in Fig~\ref{theofigSystem}. Moreover, the amplitude of $\beta_{1,j} = \beta_{2,j}$ characterizes the amplitude of the local oscillator in Fig.~1. Notably, Eq.~\eqref{eq:physicalTheorticalInitCondition} accommodates a broader class of initial conditions with $\beta_{1,j} \neq \beta_{2,j}$, which will be considered in the next section.

\subsection{Derivation of the Generalized Master Equation}

We are interested in the statistics of the measured photon numbers $n_{l,j}$ related to the photonic occupation operators $\hat b_{l,j}^\dagger \hat b_{l,j} $, i.e., unraveled according to the time index $j$ and the detector $l=1,2$.  The moment-generating function  is an alternative way to  encode the complete information about the measurement statistics of the measurement circuit shown in Fig.~\ref{theofigSystem}, and can be expressed as 
\begin{eqnarray}
M_{\boldsymbol \chi }(t)  &\equiv&  \text{tr} \left[ e^{-i \boldsymbol \chi \cdot \hat{\boldsymbol n}  } \hat U ( t ) \hat \rho(0) \hat U^\dagger ( t )  \right],
\end{eqnarray}
where $\hat U ( t )$ is the time-evolution operator corresponding to the full Hamiltonian and $\hat\rho(0)$ is the corresponding initial density matrix. We have also introduced 
\begin{equation}
\boldsymbol \chi \cdot \hat{\boldsymbol n}  =  \sum_{l = 1,2;j} \chi_{j,l} \hat b_{l,j}^\dagger \hat b_{l,j},
\end{equation}
where $\boldsymbol \chi$ is a vector with entries $\chi_{l,j} $, and $\hat{\boldsymbol n}$ is a vector with likewise ordered entries $\hat n_{l,j} = \hat b_{l,j}^\dagger \hat b_{l,j}$.  For time $t=0$, it is not hard  to show that the moment-generating function is given by
\begin{eqnarray}
M_{\boldsymbol  \chi}(0)  =  \prod_{l=1,2;j}   e^{\left(e^{-i\chi_{l , j }  } -1 \right)\left|\beta_{l , j } \right|^2},
\end{eqnarray}
which represents a Poisson statistics.

The strategy is to derive a generalized master equation  for the counting-field-dependent reduced density matrix
\begin{eqnarray}
	\hat \rho_{\boldsymbol \chi}(t) \equiv \text{tr}_{T} \left[ e^{-i \boldsymbol \chi_t \cdot \boldsymbol n  } \hat U ( t )  \hat \rho(0) \hat U^\dagger ( t ) \right],
\end{eqnarray}
where $\text{tr}_{T} \left[ \bullet \right]$ refers to the trace over the degrees of freedom of the transmission lines. Moreover, we  introduced the notation
\begin{equation}
\boldsymbol \chi_t \cdot \hat{\boldsymbol n}  =  \sum_{l = 1,2;t_j\leq t} \chi_{j,l} \hat b_{l,j}^\dagger \hat b_{l,j},
\end{equation}
incorporating only the counting fields up to time $t$.  To this end, we  express the evolution operator of the Hamiltonian in the interaction picture in Eq.~\eqref{eq:ham:rotatingWave} as
\begin{eqnarray}
 \hat U (t) &=&  \prod_{j} \hat U_j , \nonumber \\
 \hat U_j &=&  1 -i    \left(  G_j(t_j) \hat a  \hat b_{\text{R},j}^\dagger e^{-i \omega_\Delta  t_j}   +  G_j^*(t_j) \hat a^\dagger \hat b_{\text{R},j}   e^{i \omega_\Delta t_j}    \right)        dt\nonumber\\
 &&\quad-  \frac{1}{2}  \left(  G_j(t_j) \hat a  \hat b_{\text{R},j}^\dagger e^{-i \omega_\Delta  t_j}   +  G_j^*(t_j) \hat a^\dagger \hat b_{\text{R},j}   e^{i \omega_\Delta t_j}    \right)^2   dt^2 +\mathcal O(dt^2),
\end{eqnarray}
where $t_j$ is the time when the photonic modes $\hat b_{l,j}$ interacts with the quantum system. At this instant of time, the full density matrix  $\rho(t_j)$ evolves according to
\begin{eqnarray}
\hat U_j   \hat \rho    \hat U_j^\dagger   &=&  \hat \rho     \nonumber\\
&-&i dt   g_j\left(  \hat a  \hat b_{\text{R},j}^\dagger e^{-i \omega_\Delta  t}   +  \hat a^\dagger \hat b_{\text{R},j}   e^{i \omega_\Delta t}    \right)     \hat\rho    \nonumber   \\ 
&+& idt g_j \hat \rho     \left( \hat a  \hat b_{\text{R},j}^\dagger e^{-i \omega_\Delta  t}   +  \hat a^\dagger \hat b_{\text{R},j}   e^{i \omega_\Delta t}    \right)    \nonumber  \\
&+& dt^2 g_j^2  \left(  \hat a  \hat b_{\text{R},j}^\dagger e^{-i \omega_\Delta  t}   +  \hat a^\dagger \hat b_{\text{R},j}   e^{i \omega_\Delta t}    \right)         \hat\rho   \left(   \hat a  \hat b_{\text{R},j}^\dagger e^{-i \omega_\Delta  t}   +  \hat a^\dagger \hat b_{\text{R},j}   e^{i \omega_\Delta t}    \right)   \nonumber   \\
&-& dt^2   \frac{1}{2}  g_j^2    \left \lbrace \left(  \hat a  \hat b_{\text{R},j}^\dagger e^{-i \omega_\Delta  t}   +  \hat a^\dagger \hat b_{\text{R},j}   e^{i \omega_\Delta t}    \right)^2     , \hat \rho   \right \rbrace,
\end{eqnarray}
where we have neglected the time arguments for brevity. Using Eq.~\eqref{eq:measurementBasisTransformation}, we incorporate the squeezing and represent the time evolution in the measurement basis. To obtain the time evolution of the corresponding reduced density matrix, we evaluate
\begin{equation}
\hat\rho_{\boldsymbol\chi}(t_j+dt)  =   \text{tr}_{j}  \left[  e^{ -i \sum_{l}\chi_{j,l}   \hat n_{j,l} } \hat U_j   \hat\rho_{\boldsymbol \chi}(t_j)\otimes \prod_{l} \left| \beta_{j,l} \right>\left< \beta_{j,l}  \right|    \hat U_j^\dagger   \right]  ,
\end{equation}
where $\text{tr}_j$ refers to the trace over the Fock states of $\hat b_{l,j}$ with $l=1,2$. Using the identity,
\begin{eqnarray}
 \hat b_{l,j}e^{-i \chi_{l,j}   \hat n_{l,j}    }   =e^{-i \chi_{l,j}   \hat n_{l,j}   } \hat b_{l,j}e^{- i\chi_{l , j } },\nonumber\\
  e^{-i \chi_{l,j}   \hat n_{l,j}    }\hat b_{l,j}^\dagger   = \hat b_{l,j}^\dagger e^{- i\chi_{l , j } } e^{-i \chi_{l,j}   \hat n_{l,j}   }     ,
\end{eqnarray}
we find
\begin{eqnarray}   
\hat\rho_{\boldsymbol \chi}(t+dt)  &=&    \hat\rho_{\boldsymbol \chi}  M_{\boldsymbol  \chi,j}   \nonumber\\
&-&i dt g_j  \left( u  \beta_{1,j} -   u  \beta_{2,j}   - v  \beta_{1,j}^* e^{-i\chi_{1,j}}  +  v \beta_{2,j}^*e^{-i\chi_{2,j}} \right)  \hat  a^\dagger e^{i \omega_\Delta  t} \hat\rho_{\boldsymbol \chi} M_{\boldsymbol  \chi,j}  \nonumber      \\   
&-&i dt   g_j      \left( u  \beta_{1,j}^*e^{-i\chi_{1,j}}   -   u  \beta_{2,j}^*e^{-i\chi_{2,j}}    - v  \beta_{1,j}  +  v \beta_{2,j}\right)  \hat  a e^{-i \omega_\Delta  t} \hat\rho_{\boldsymbol \chi} M_{\boldsymbol  \chi,j}  \nonumber      \\ 
&+&i dt   g_j   \left( u  \beta_{1,j} e^{-i\chi_{1,j}}  -   u  \beta_{2,j}e^{-i\chi_{2,j}}   - v  \beta_{1,j}^* +  v \beta_{2,j}^* \right)     \hat\rho_{\boldsymbol \chi}  \hat  a^\dagger e^{i \omega_\Delta  t}   M_{\boldsymbol  \chi,j}  \nonumber    \\ 
&+&i dt   g_j  \left( u  \beta_{1,j}^*  -   u  \beta_{2,j}^*   - v  \beta_{1,j} e^{-i\chi_{1,j}}  +  v \beta_{2,j}e^{-i\chi_{2,j}}    \right) \hat\rho_{\boldsymbol \chi}  \hat  a  e^{-i \omega_\Delta  t}    M_{\boldsymbol  \chi,\alpha,j}    \nonumber     \\ 
&+& dt^2      g_j^2   u^2  \left(e^{-i\chi_{1,j}}   +   e^{-i\chi_{2,j}} \right)  \hat  a   \hat\rho_{\boldsymbol \chi}  \hat  a^\dagger    M_{\boldsymbol  \chi,j}   \nonumber \\
&+& dt^2        g_j^2  v^2  \left( e^{-i\chi_{1,j}}   +    e^{-i\chi_{2,j}} \right)    \hat  a^\dagger   \hat \rho_{\boldsymbol \chi}  \hat  a    M_{\boldsymbol  \chi,j}  \nonumber \\
&-& dt^2       g_j^2 uv \left( e^{-i\chi_{1,j}}   +   e^{-i\chi_{2,j}} \right)     \hat  a_d    \hat\rho_{\boldsymbol \chi}  \hat  a_d   e^{-2i \omega_\Delta  t}  M_{\boldsymbol  \chi,j}   \nonumber \\
&-& dt^2      g_j^{*2}  uv  \left(  e^{-i\chi_{1,j}}   +    e^{-i\chi_{2,j}} \right)    \hat  a_d^\dagger    \hat\rho_{\boldsymbol \chi}  \hat  a_d^\dagger  e^{2i \omega_\Delta  t}   M_{\boldsymbol  \chi,j}   \nonumber \\
&-& dt^2     g_j^2  u^2    \left \lbrace \hat  a^\dagger      \hat  a, \hat\rho_{\boldsymbol \chi}   \right \rbrace  M_{\boldsymbol  \chi,j}  \nonumber  \\
&-& dt^2      g_j^2   v^2    \left \lbrace \hat  a    \hat  a^\dagger  , \hat\rho_{\boldsymbol \chi}   \right \rbrace  M_{\boldsymbol  \chi,j}   \nonumber \\
&+& dt^2      g_j^2   u v      \left \lbrace \hat  a   \hat  a, \hat\rho_{\boldsymbol \chi}   \right \rbrace e^{-2i \omega_\Delta  t} M_{\boldsymbol  \chi,j}  \nonumber  \\
&+& dt^2      g_j^{2} u v    \left \lbrace \hat  a^\dagger      \hat  a^\dagger  , \hat\rho_{\boldsymbol \chi}   \right \rbrace   e^{2i \omega_\Delta  t}   M_{\boldsymbol  \chi,j}   \nonumber \\
&+& \mathcal O(dt^2),
\label{eq:densityMatrtix:timeIncrement}
\end{eqnarray}
where 
\begin{eqnarray}
M_{\boldsymbol  \chi,l,j}  &=&     e^{\left(e^{-i\chi_{l , j }  } -1 \right)\left|\beta_{l , j } \right|^2}\nonumber  \\
  &\approx&   1+ \left(e^{-i\chi_{l , j }  } -1 \right)  \left|\beta_{l , j } \right|^2 +\mathcal O \left( dt^2\right)
\end{eqnarray}
is the moment-generating function of $\hat n_{l,j}$ before interacting with the resonator. Recalling that $\beta_{l , j } \propto \sqrt{dr/c} \propto \sqrt{ dt} $, we have expanded the exponential up to first order in $dt$ in the second line.

To proceed, we assume a time-translational-invariant system such that the parameters in Eq.~\eqref{eq:densityMatrtix:timeIncrement} can be replaced by
\begin{eqnarray}
	g_j & =& \frac{\tilde g}{\sqrt{dt}} , \nonumber \\
	\beta_{l,j} &=&  \beta_{l} \sqrt{dt},
\end{eqnarray}
where we also explicitly express the scaling with the time increment $dt$. Moreover, we consider the cumulative counting of photons by replacing $\chi_{l,j} =\chi_{l}$. Evaluating Eq.~\eqref{eq:densityMatrtix:timeIncrement}, we obtain the generalized master equation
\begin{eqnarray}
	\frac{d}{dt}\hat \rho_{\boldsymbol  \chi}(t) =   &-& i \left[\omega_\Delta \hat a^\dagger \hat a ,\hat\rho_{\boldsymbol  \chi} \right] \nonumber \\
	 &-&i   \tilde g   \left( u  \beta_{1} -  u  \beta_{2}   - v  \beta_{1}^* e^{-i\chi_{1}}  +  v  \beta_{2}^*e^{-i\chi_{2}} \right)  \hat  a^\dagger   \hat\rho_{\boldsymbol  \chi}  \nonumber    \\ 
	&-&i   \tilde g       \left( u  \beta_{1}^*e^{-i\chi_{1}}  -  u  \beta_{2}^* e^{-i\chi_{2}}   - v  \beta_{1}  +  v  \beta_{2} \right) \hat  a \hat \rho_{\boldsymbol  \chi}  \nonumber      \\ 
	&+&i   \tilde g   \left( u  \beta_{1}e^{- i\chi_{1}}   -  u  \beta_{2}e^{- i\chi_{2}}   - v  \beta_{1}^*  +  v  \beta_{2}^* \right)   \hat\rho_{\boldsymbol  \chi}  \hat  a^\dagger   \nonumber    \\ 
	&+&i  \tilde g    \left( u  \beta_{1}^* -  u  \beta_{2}^*    - v  \beta_{1}  e^{ -i\chi_{1}} +  v  \beta_{2} e^{- i\chi_{2}} \right) \hat \rho_{\boldsymbol  \chi}  \hat  a      \nonumber     \\ 
	&+&      \tilde g^2   u^2  \left(e^{-i\chi_{1}} + e^{-i\chi_{2}} \right)  \hat  a   \hat\rho_{\boldsymbol  \chi}  \hat  a^\dagger   
	+     \tilde g ^2   v^2  \left(e^{-i\chi_{1}} + e^{-i\chi_{2}} \right)   \hat  a^\dagger   \hat \rho_{\boldsymbol  \chi}  \hat  a       \nonumber \\
	&-&      \tilde g^{2}   u v \left(e^{-i\chi_{1}} + e^{-i\chi_{2}} \right)   \hat  a    \hat\rho_{\boldsymbol  \chi}  \hat  a -      \tilde g^{2 }  u v \left(e^{-i\chi_{1}} + e^{-i\chi_{2}} \right)   \hat  a^\dagger    \hat\rho_{\boldsymbol  \chi}  \hat  a^\dagger      \nonumber \\
	&-&    \tilde g ^2 u^2    \left \lbrace \hat  a^\dagger      \hat  a,\hat \rho_{\boldsymbol  \chi}   \right \rbrace   -      \tilde g  ^2 v^2    \left \lbrace \hat  a    \hat  a^\dagger  , \hat\rho_{\boldsymbol  \chi}   \right \rbrace    
	+      \tilde g^{2}  u v      \left \lbrace \hat  a  \hat  a, \hat\rho_{\boldsymbol  \chi}   \right \rbrace 
	+     \tilde g^{2}  u v    \left \lbrace \hat  a^\dagger      \hat  a^\dagger  , \hat\rho_{\boldsymbol  \chi}   \right \rbrace \nonumber \\
	&+&  \left|\beta_{1}\right|^2  \left(e^{-i\chi_{1}}  -1  \right) \hat\rho_{\boldsymbol  \chi} + \left|\beta_{2}\right|^2\left(e^{-i\chi_{2}}  -1  \right) \hat\rho_{\boldsymbol  \chi}.
\end{eqnarray}

This can also be expressed in a more compact form 
\begin{eqnarray}
\frac{d}{dt}\hat\rho_{\boldsymbol  \chi}(t)  =  &-& i \left[\omega_\Delta \hat a^\dagger \hat a ,\hat\rho_{\boldsymbol  \chi} \right] \nonumber \\
&-&  \frac{i}{2} \left[ \tilde g u \beta_{1} \hat a^\dagger  + \tilde g  u \beta_{1}^* \hat a  - \tilde g v  \beta_{1}^* \hat a^\dagger  - \gamma v  \beta_{1} \hat a  ,\hat\rho_{\boldsymbol  \chi} \right]    \nonumber  \\
&+&  \frac{i}{2} \left[ \tilde g u \beta_{2} \hat a^\dagger  + \tilde g u \beta_{2}^* \hat a  - \tilde g v\beta_{2}^* \hat a^\dagger  -\tilde g v \beta_{2} \hat a  ,\hat\rho_{\boldsymbol  \chi} \right]   \nonumber \\
&+&  D_{\chi_1} \left[ \tilde g u\hat a     -  \tilde g v\hat a^\dagger   + i \beta_{1} \right] \hat\rho_{\boldsymbol  \chi}   \nonumber \\ 
&+&  D_{\chi_2} \left[ \tilde g u\hat a -  \tilde g v\hat a^\dagger  - i \beta_{2}^*   \right] \hat\rho_{\boldsymbol  \chi}  ,
\end{eqnarray}
where the counting-field-dependent dissipator is defined as
\begin{equation}
    D_{\chi} \left[A \right]\rho  = e^{-i\chi}\hat A \rho \hat A^\dagger - \frac{1}{2} \left\lbrace  \rho ,\hat A^\dagger \hat A  \right \rbrace,
\end{equation}
for an arbitrary system operator $\hat A$.

Finally, we add a driving term imposed by the transmission line $l =\text{L}$ along similar lines as for the transmission line $l=\text{R}$. In doing so, we obtain
\begin{eqnarray}
\frac{d}{dt}\hat\rho_{\boldsymbol  \chi}(t)  =  &-& i \left[\omega_\Delta \hat a^\dagger \hat a ,\hat\rho_{\boldsymbol  \chi} \right] \nonumber \\
&-&  \frac{i}{2} \left[ \tilde g u \beta_{1} \hat a^\dagger  + \tilde g u \beta_{1}^* \hat a  - \tilde g v  \beta_{1}^* \hat a^\dagger  - \tilde g v  \beta_{1} \hat a  ,\hat\rho_{\boldsymbol  \chi} \right]    \nonumber  \\
&+&  \frac{i}{2} \left[ \tilde g u \beta_{2} \hat a^\dagger  + \tilde g u \beta_{2}^* \hat a  - \tilde gv\beta_{2}^* \hat a^\dagger  - \tilde gv \beta_{2} \hat a  ,\hat\rho_{\boldsymbol  \chi} \right]   \nonumber \\
&+&  i \left[ \tilde g_{\text{L}}\beta_{\text{L}} \hat a^\dagger  + \tilde g_{\text{L}}  \beta_{\text{L}}^* \hat a   ,\hat\rho_{\boldsymbol  \chi} \right] \nonumber \\
&+&  D_{\chi_1} \left[ \tilde g u\hat a     -  \tilde g v\hat a^\dagger   + i \beta_{1} \right] \hat\rho_{\boldsymbol  \chi}   \nonumber \\ 
&+&  D_{\chi_2} \left[ \tilde g u\hat a -  \tilde g v\hat a^\dagger  - i \beta_{2}^*   \right] \hat\rho_{\boldsymbol  \chi}  \nonumber \\ 
&+&  D\left[ \tilde g_{\text{L}}  \hat a   \right] \hat\rho_{\boldsymbol  \chi} ,
\label{eq:amendedMasterEquation}
\end{eqnarray}
where $\beta_{\text{L}}$ denotes the amplitude of the driving field, $\tilde g_{\text{L}}  =g_{\text{L}} \sqrt{dt}$ is the effective coupling between system and transmission line. To prevent information leaking into the transmission line $l=\text{L}$, experiments typically use $\tilde g_{\text{L}} \rightarrow 0$ and a large driving amplitude $\left| \beta_{\text{L}} \right|\rightarrow \infty$ such that $\tilde g_{\text{L}}\beta_{\text{L}}  =-i\Omega $ is constant. In this limit, we find that the generalized master equation takes the form of 
\begin{eqnarray}
\frac{d}{dt}\hat\rho_{\boldsymbol  \chi}  =  &-& i \left[\omega_\Delta \hat a^\dagger \hat a  -i \Omega\hat a^\dagger +i\Omega^*\hat a  ,\hat\rho_{\boldsymbol  \chi} \right] \nonumber \\
&+&  \frac{i}{2\sqrt{2}} \sum_{l=1,2}(-1)^l\left[ \left( \tilde g  u \beta_{l} -\tilde g v  \beta_{l}^*  \right) \hat a^\dagger  + \left( \tilde g  u \beta_{l}^* -    \tilde g v  \beta_{l} \right)\hat a ,\hat\rho_{\boldsymbol  \chi}   \right]    \nonumber  \\
&+&  \frac{1}{2} D_{\chi_1} \left[\tilde g u\hat a     -  \tilde g v\hat a^\dagger   + i \sqrt{2}\beta_{1} \right] \hat\rho_{\boldsymbol  \chi}   \nonumber \\ 
&+&  \frac{1}{2} D_{\chi_2} \left[ \tilde g u\hat a -  \tilde gv\hat a^\dagger  - i \sqrt{2}\beta_{2}^*   \right]\hat \rho_{\boldsymbol  \chi} ,  
\end{eqnarray}
which is the same master equation as Eq.~(4) in the Letter for real $\beta_1 =\beta_2$ and $\gamma = \tilde g^2$.

\section{Non-Hermitian mean-field approach}

In this section, we introduce a generalized mean-field approach, which transforms the generalized master equation in a non-Hermitian fashion, such that we can identify the leading terms of the cumulant-generating function and the quantum Fisher information.

\subsection{Non-Hermitian mean-field  expansion of the cumulant-generating function}

\label{app:cumulantGenFunction}

Our starting point is the generalized master equation describing the dispersive readout system
\begin{eqnarray}
\dot{\hat{\rho}}_{\boldsymbol \chi} =-i\left[\omega_\Delta \hat{a}^{\dagger} \hat{a}-i\Omega \hat{a}^{\dagger}+i\Omega \hat{a}, \hat{\rho}_{\boldsymbol{\chi}}\right]+\frac{1}{2}D_{\chi_1}\left[\sqrt{\gamma}\mu\hat{a} -\sqrt{\gamma} v \hat{a}^{\dagger}-i \beta\right] \hat{\rho}_{\boldsymbol{\chi}} +\frac{1}{2} D_{\chi_{2}}\left[\sqrt{\gamma}\mu\hat{a} -\sqrt{\gamma} v \hat{a}^\dagger+i \beta\right] \hat{\rho}_{\boldsymbol{\chi}} ,
\label{eq:generalizedMasterEquation}
\end{eqnarray}
which determines the moment-generating function of the measured photonic probability distribution via $M_{\boldsymbol \chi} =\text{Tr} \left[ \hat{\rho}_{\boldsymbol \chi} \right] $. For brevity of notation, we restrict the explanations to the linear system, but remark that the generalization to non-linear system is straightforward. 
To identify the leading-order contributions of the moment-generating function in terms of a mean-field approach,  we define the non-unitary displacement operators
\begin{eqnarray}
D^{-1}\left(\alpha_{\text{f}}, \alpha_{\text{f}}^{+}\right) &=& \exp\left[{\alpha_{\text{f}} \hat a^{\dagger} + \alpha_{\text{f}}^+ \hat a}\right],\nonumber \\
\qquad D^{-1}\left(\alpha_{\text{b}}, \alpha_{\text{b}}^{+}\right) &=& \exp\left[{\alpha_{\text{b}} \hat a^{\dagger} + \alpha_{\text{b}}^+ \hat a}\right],
\end{eqnarray}
where $\alpha_{\text{f}}$, $\alpha_{\text{f}}^+$, $\alpha_{\text{b}}$, $\alpha_{\text{b}}^+$ are four distinct undetermined complex numbers.  Applied to the photonic creation and annihilation operators, we find
\begin{eqnarray}
D^{-1}\left(\alpha_{\text{f}}, \alpha_{\text{f}}^{+}\right) \hat{a} D\left(\alpha_{\text{f}}, \alpha_{\text{f}}^{+}\right) &=& \hat a +\alpha_{\text{f}} \nonumber, \\
\qquad D^{-1}\left(\alpha_{\text{f}}, \alpha_{\text{f}}^{+}\right) \hat{a}^\dagger 
D\left(\alpha_{\text{f}}, \alpha_{\text{f}}^{+}\right) &=& \hat a^\dagger +\alpha_{\text{f}}^+\nonumber,\\
D^{-1}\left(\alpha_{\text{b}}, \alpha_{\text{b}}^{+}\right) \hat{a} D\left(\alpha_{\text{b}}, \alpha_{\text{b}}^{+}\right) &=& \hat a +\alpha_{\text{b}} \nonumber, \\
D^{-1}\left(\alpha_{\text{b}}, \alpha_{\text{b}}^{+}\right) \hat{a}^\dagger D\left(\alpha_{\text{b}}, \alpha_{\text{b}}^{+}\right) &=& \hat a^\dagger +\alpha_{\text{b}}^+,
\end{eqnarray}
such that the transformed creation operators are not the complex conjugates of the transformed annihilation operators. Consequently, the transformed reduced density matrix  $\hat{R}_{\boldsymbol{\chi}}  \equiv D^{-1}\left(\alpha_{\text{f}}, \alpha_{\text{f}}^{+}\right) \hat{\rho}_{\boldsymbol{\chi}} D\left(\alpha_{\text{b}}, \alpha_{\text{b}}^{+}\right)$, which is not Hermitian anymore, is governed by the transformed master equation
\begin{eqnarray}
\dot{\hat{R}}_{\boldsymbol{\chi}}&=&\mathcal K_{\boldsymbol \chi} \hat{R}_{\boldsymbol{\chi}}  + {\cal L}^{(1)}\left[\hat a \right] \hat{R}_{\boldsymbol{\chi}} +{\cal L}^{(2)}\left[\hat a \right] \hat{R}_{\boldsymbol{\chi}},
\label{eq:approximateMasterEquation}
\end{eqnarray}
where the first term is just a multiplaction of the generalized density matrix by the scalar function 
\begin{eqnarray}
\mathcal K_{\boldsymbol{\chi}} =&& -i\left(\omega_\Delta \alpha_{\text{f}}^{+} \alpha_{\text{f}}-i\Omega \alpha_{\text{f}}^{+}+ i\Omega \alpha_{\text{f}}\right)  +i \left(\omega_\Delta \alpha_{\text{b}} ^{+} \alpha_{\text{b}} -i\Omega \alpha_{\text{b}} ^{+}+i\Omega \alpha_{\text{b}} \right) \nonumber\\
&&+\frac{\gamma}{2} \left[ \left( e^{-i \chi_1} + e^{-i \chi_2} \right) \xi_{\text{f} }\eta_{\text{b}}  -\eta_{\text{f}}\xi_{\text{f}}  -\eta_{ \text{b}} \xi_{\text{b}}  \right] \nonumber\\
&& +\frac{\sqrt{\gamma}}{2}  \left( e^{-i \chi_1} - e^{-i \chi_2} \right) \left( i \xi_{\text{f} } \beta^* -i  \beta \eta_{\text{b}} \right)    \nonumber\\
&& +\left| \beta\right|^2\left( e^{-i \chi_1} + e^{-i \chi_2} -2 \right)
\label{scalar_Function_F}
\end{eqnarray}
 with the coefficients defined by
$ \xi_{\text{f} } = \mu \alpha_{\text{f} }-\nu \alpha_{\text{f} } ^{+}$, 
$ \xi_{\text{b}} = \mu \alpha_{\text{b}} - \nu \alpha_{\text{b}}^{+}$, $\eta_{\text{f} } = \mu \alpha_{\text{f} }^{+}-\nu \alpha_{\text{f} },$ and 
$\eta_{\text{b}} = \mu \alpha_{\text{b}}^{+} - \nu \alpha_{\text{b}}$.
 The super-operators ${\cal L}^{(1)}_{\boldsymbol \chi}[\hat a]$ and ${\cal L}^{(2)}_{\boldsymbol \chi}[\hat a]$ characterize first- and second-order contributions in the operators $\hat a$ and $\hat a^\dagger$, respectively, and are given by
\begin{eqnarray}
{\cal L}^{(1)}_{\boldsymbol \chi}\left[\hat a \right] \hat{R}_{\boldsymbol{\chi}}&=&A^+\hat{a} \hat{R}_{\boldsymbol{\chi}} + B\hat{R}_{\boldsymbol{\chi}} \hat{a}^{\dagger} + A\hat{a}^{\dagger} \hat{R}_{\boldsymbol{\chi}} + B^+ \hat{R}_{\boldsymbol{\chi}} \hat{a},\nonumber\\
{\cal L}_{\chi}^{(2)}\left[\hat a \right]\hat{R}_{\boldsymbol{\chi}}&=&-i\left[\omega_\Delta \hat{a}^{\dagger} \hat{a}, \hat{R}_{\boldsymbol{\chi}}\right]+D_{\chi_1}\left[\mu\hat{a} -\nu \hat{a}^{\dagger}\right] \hat{R}_{\boldsymbol{\chi}} +D_{\chi_{2}}\left[\mu\hat{a} -\nu \hat{a}^\dagger \right] \hat{R}_{\boldsymbol{\chi}},
\end{eqnarray}
with coefficients
\begin{eqnarray}
A &=& -i\left(\omega_\Delta \alpha_{\text{f}} -i\Omega \right)  - \frac{\gamma }{2}\left( e^{-i \chi_1} + e^{-i \chi_2} \right)\nu \eta_{\text{b}}  -\frac{\gamma}{2} \left(\mu \xi_{\text{f}}  - \nu \eta_{\text{f}}  \right) \nonumber \\
&&-\frac{\sqrt{\gamma} }{2} \left( e^{-i \chi_1} - e^{-i \chi_2} \right)  i \nu \beta^*  ,\nonumber\\
A^+ &=& -i\left(\omega_\Delta \alpha_{\text{f}}^+ +i\Omega \right)  + \frac{ \gamma }{2}\left( e^{-i \chi_1} + e^{-i \chi_2} \right)\mu \eta_{\text{b}}  -\frac{\gamma}{2} \left(-\nu \xi_{\text{f}}  + \mu \eta_{\text{f}}  \right) \nonumber \\
&&+\frac{\sqrt{\gamma}} {2} \left( e^{-i \chi_1} - e^{-i \chi_2} \right)  i \mu \beta^*  ,\nonumber\\
B &=& i\left(\omega_\Delta \alpha_{\text{b}} -i\Omega \right)  + \frac{\gamma}{2} \left( e^{-i \chi_1} + e^{-i \chi_2} \right)\mu \xi_{\text{f}}  -\frac{\gamma}{2} \left(\mu \xi_{\text{b}}  - \nu \eta_{\text{b}}  \right) \nonumber \\
&&-\frac{\sqrt{\gamma}}{2}  \left( e^{-i \chi_1} - e^{-i \chi_2} \right)  i \mu \beta  ,\nonumber\\
B^+ &=& i\left(\omega_\Delta \alpha_{\text{b}}^+ +i\Omega \right)  + \frac{\gamma}{2} \left( e^{-i \chi_1} + e^{-i \chi_2} \right)\nu \xi_{\text{f}}  -\frac{\gamma}{2} \left(-\nu \xi_{\text{b}}  + \mu \eta_{\text{b}}  \right) \nonumber \\
&&+ \frac{\sqrt{\gamma}}{2}  \left( e^{-i \chi_1} - e^{-i \chi_2} \right)  i \nu \beta  .\nonumber\\
\end{eqnarray}
To determine the complex mean fields $(\alpha_{\text{f}},\alpha_{\text{f}}^+,\alpha_{\text{b}},\alpha_{\text{b}}^+)$ we set $A=A^+=B=B^+ = 0$, and solve for $(\alpha_{\text{f}},\alpha_{\text{f}}^+,\alpha_{\text{b}},\alpha_{\text{b}}^+)$. For this linear system, an analytical solution can be found using a computer algebra system like Mathematica. For non-linear systems, this calculation must be carried out numerically. In doing so, the linear contribution of the generalized master equation vanishes, ${\cal L}^{(1)}\left[\hat a \right] \hat{R}_{\boldsymbol{\chi}}=0$. Furthermore, we neglect the quadratic terms ${\cal L}^{(2)}\left[\hat a \right] \hat{R}_{\boldsymbol{\chi}}\rightarrow 0$. The resulting approximated master equation in Eq.~\eqref{eq:approximateMasterEquation} has the solution
\begin{equation}
    \hat{R}_{\boldsymbol{\chi}}(t)  = e^{\mathcal K_{\boldsymbol \chi} t} D\left(\alpha_{\text{f}}, \alpha_{\text{f}}^{+}\right)^{-1}\rho(0)D\left(\alpha_{\text{b}}, \alpha_{\text{b}}^{+}\right),
\end{equation}
which in the original picture trivially becomes
\begin{equation}
    \hat{\rho}_{\boldsymbol{\chi}}(t)  = e^{\mathcal K_{\boldsymbol \chi} t} \rho(0),
\end{equation}
such that the mean-field contribution of the cumulant-generating function is given by
\begin{eqnarray}
K_{\boldsymbol \chi}(t) &=& \ln\left[{M}_{\boldsymbol \chi} (t)\right] = \mathcal  K_{\boldsymbol \chi} t.
\end{eqnarray}
The cumulants can now be obtained by deriving this cumulant-generating function with respect to the counting fields. As the calculation is straightforward but tedious, we employ the computer algebra system Mathematica to find the lengthy expressions for the first four cumulants, which are plotted in Fig. 2 of the Letter. As explained in the Letter, the cumulative Fisher information in the asymptotic limit plotted in Figs. 1 and Fig. 3 can be obtained using the first two cumulants, such that we obtain Eq.~(11) in the Letter.

\subsection{Quantum Fisher information}

\label{sec:fisherInfo:derivation}

In terms of the quantum state $\left| \phi \right> $ of the total system (i.e.,  resonator and transmission lines), the quantum Fisher information for estimation of a system parameter (here: $\omega_{\Delta}$)  can be  expressed as~\cite{zanardi2007information},
\begin{equation}
    \mathcal{I}^{Q}_{\omega_{\Delta}}  = 4\left( \left< \partial_{\omega_\Delta}\phi \mid \partial_{\omega_\Delta} \phi\right>  - \left| \left< \phi \mid \partial_{\omega_\Delta} \phi\right>\right|^2\right) .
\end{equation}
When dealing with open quantum systems, this can be written as
\begin{equation}
\mathcal{I}^{Q}_{\omega_{\Delta}}  = -\partial^{2}_{\delta} \log \mathrm{Tr}[\hat\rho_\delta] \bigg|_{\delta=0},
\label{FIX_def}
\end{equation}
where the $\delta$-dependent density matrix evolves according to
\begin{eqnarray}
\dot{\hat{\rho}}_{ \delta} =-i\left[\omega_{\Delta_f} \hat{a}^{\dagger} \hat{a}\hat{\rho}_{{\delta}}-\omega_{\Delta_b}\hat{\rho}_{\delta} \hat{a}^{\dagger} \hat{a}\right]-i\left[-i\Omega \hat{a}^{\dagger}+i\Omega \hat{a}, \hat{\rho}_\delta\right]+D\left[\sqrt{\gamma}\mu\hat{a} -\sqrt{\gamma}v \hat{a}^{\dagger}
\right] \hat{\rho}_{\delta},
\end{eqnarray}
with $\omega_{\Delta_f}=\omega_\Delta+\delta$ and $\omega_{\Delta_b}=\omega_\Delta-\delta$~\cite{gammelmark2014fisher}. This is a generalized master equation reminiscent of the one in Eq.~\eqref{eq:generalizedMasterEquation} determining the moment-generating function. To calculate the quantum Fisher information, we can thus apply the same procedures as for the cumulants.

 Performing the non-Hermitian mean-field expansion, we obtain 
\begin{eqnarray}
\dot{\hat R}_\delta&=& F_\delta\hat{R}_\delta  +  {\cal L}^{(1)}_\delta[ \hat a] \hat{R}_\delta +{\cal L}^{(2)}_\delta [ \hat a]  \hat{R}_\delta,
\end{eqnarray}
where the scalar function is given by
\begin{eqnarray}
F_\delta =&& -i\left(\omega_{\Delta_f} \alpha_{\text{f}}^{+} \alpha_{\text{f}}-i\Omega \alpha_{\text{f}}^{+}+ i\Omega \alpha_{\text{f}}\right)  +i \left(\omega_{\Delta_b} \alpha_{\text{b}} ^{+} \alpha_{\text{b}} -i\Omega \alpha_{\text{b}} ^{+}+i\Omega \alpha_{\text{b}} \right) \nonumber\\
&&+\gamma \left[ \xi_{\text{f} }\eta_{\text{b}}  -\frac{1}{2} \eta_{\text{f}}\xi_{\text{f}}  -\frac{1}{2} \eta_{ \text{b}} \xi_{\text{b}}  \right] 
\end{eqnarray}
 with the coefficients $ \xi_{\text{f} } 
$, $ \xi_{\text{b}}$, $\eta_{\text{f} }$, and 
$\eta_{\text{b}}$ given below Eq.~\eqref{scalar_Function_F}. The  super-operators ${\cal L}^{(1)}_\delta[\hat a]$ and ${\cal L}^{(2)}_\delta[\hat a]$ characterize first- and second-order contributions of the operators $\hat a$ and $\hat a^\dagger$, respectively, and read
\begin{eqnarray} 
{\cal L}_\delta^{(1)} [ \hat a]  \hat{R}_{\delta}&=&A^+\hat{a} \hat{R}_{\delta} + B\hat{R}_{\delta} \hat{a}^{\dagger} + A\hat{a}^{\dagger} \hat{R}_{\delta} + B^+ \hat{R}_{\delta} \hat{a},\nonumber\\
{\cal L}_\delta^{(2)} [ \hat a]  \hat{R}_{\delta}&=&-i\left[\omega_{\Delta_l} \hat{a}^{\dagger} \hat{a}\hat{R}_{{\delta}}-\omega_{\Delta_r}\hat{R}_{\delta} \hat{a}^{\dagger} \hat{a}\right]+D \left[\mu\hat{a} -\nu \hat{a}^{\dagger}\right] \hat{R}_{\delta}  ,
\end{eqnarray}
with coefficients
\begin{eqnarray}
A &=& -i\left(\omega_{\Delta_f} \alpha_{\text{f}} -i\Omega \right)  - \gamma \nu \eta_{\text{b}}  -\frac{\gamma}{2} \left(\mu \xi_{\text{f}}  - \nu \eta_{\text{f}}  \right) ,\nonumber \\
A^+ &=& -i\left(\omega_{\Delta_f} \alpha_{\text{f}}^+ +i\Omega \right)  + \gamma \mu \eta_{\text{b}}  -\frac{\gamma}{2} \left(-\nu \xi_{\text{f}}  + \mu \eta_{\text{f}}  \right) ,\nonumber \\
B &=& i\left(\omega_{\Delta_b} \alpha_{\text{b}} -i\Omega \right)  + \gamma\mu \xi_{\text{f}}  -\frac{\gamma}{2} \left(\mu \xi_{\text{b}}  - \nu \eta_{\text{b}}  \right) ,\nonumber \\
B^+ &=& i\left(\omega_{\Delta_b} \alpha_{\text{b}}^+ +i\Omega \right)  - \gamma \nu \xi_{\text{f}}  -\frac{\gamma}{2} \left(-\nu \xi_{\text{b}}  + \mu \eta_{\text{b}}  \right).
\end{eqnarray}
As in Sec.~\ref{app:cumulantGenFunction}, we set $A=A^+=B=B^+ = 0$,  and resolve for the complex mean fields, either analytically for the linear system using a computer algebra system,  or numerically for the nonlinear system. Neglecting the quadratic terms in the photonic operators, we thus obtain the approximated master equation
\begin{eqnarray}
\dot{\hat{R}}_\delta&=& F_\delta\hat{R}_\delta.
\end{eqnarray}
Following now similar lines as in Sec.~\ref{app:cumulantGenFunction}, we finally obtain the mean-field expression for the quantum Fisher information 
\begin{eqnarray}
{\cal I }_q = -\left.\frac{\partial^2 F_\delta}{\partial\delta^2}\right|_{\delta=0} t.
\end{eqnarray}
which can be effectively evaluated using the computer algebra system Mathematica, such that we obtain Eq. (14) in the Letter.

\bibliography{bibliography}